\documentclass[a4paper, 12pt]{article}

\pdfoutput = 1

\usepackage{styleBuding}
\usepackage{bm}
\usepackage{color,listings}
\usepackage[usenames,dvipsnames,svgnames,table]{xcolor}
\usepackage{amsthm, amsmath}
\usepackage{amssymb}
\usepackage{mathrsfs}
\usepackage{graphicx}
\usepackage{slashed}
\usepackage{soul}
\usepackage{multirow}
\usepackage{subfigure}
\usepackage{siunitx} % for physics units
\usepackage{url} %for link in @article
\usepackage[utf8]{inputenc}
\usepackage[figuresright]{rotating}
\definecolor{back}{HTML}{F8F8F8}
\usepackage{epstopdf}

\newcommand{\tev}{\ensuremath{\,\text{TeV}}}

\newcommand{\rom}[1]{\uppercase\expandafter{\romannumeral #1\relax}}

\allowdisplaybreaks
\DeclareUnicodeCharacter{2212}{-}

\let\jnfont=\rm
\def\NPB#1,{{\jnfont Nucl.\ Phys.\ B }{\bf #1},}
\def\PLB#1,{{\jnfont Phys.\ Lett.\ B }{\bf #1},}
\def\EPJC#1,{{\jnfont Eur.\ Phys.\ Jour.\ C }{\bf #1},}
\def\PRD#1,{{\jnfont Phys.\ Rev.\ D }{\bf #1},}
\def\PRL#1,{{\jnfont Phys.\ Rev.\ Lett.\ }{\bf #1},}
\def\MPLA#1,{{\jnfont Mod.\ Phys.\ Lett.\ A }{\bf #1},}
\def\JPG#1,{{\jnfont J.\ Phys.\ G}{\bf #1},}
\def\CTP#1,{{\jnfont Commun.\ Theor.\ Phys.\ }{\bf #1},}
\def\ZPC#1,{{\jnfont Z.\ Phys.\ C }{\bf #1},}
\def\JHEP#1,{{\jnfont JHEP \ }{\bf #1},}

\title{Impact of LHC probes of SUSY and recent measurement of $(g-2)_{\mu}$ on $\mathbb{Z}_3$-NMSSM }
\author{Junjie Cao$^{a}$, Fei Li$^a$, Jingwei Lian$^a$, Yusi Pan$^a$, Di Zhang$^a$}
\affiliation{ $^a$ Department of Physics, Henan Normal University, Xinxiang 453007, China}
\emailAdd{junjiec@alumni.itp.ac.cn}
\emailAdd{dz481655@gmail.com}
\abstract{It is well known that excessively heavy supersymmetric particles (sparticles) are disfavored to explain the $(g-2)_\mu$ anomaly, but some people overlook that moderately light sparticles are also disfavored by the LHC probes of supersymmetry. We take the Next-to-Minimal Supersymmetric Standard Model as an example to emphasize the latter point. It is found that, if the theory is required to explain the anomaly at $2\sigma$ level and meanwhile keep consistent with the LHC results, the following lower bounds may be set: $\tan \beta \gtrsim 20$, $|M_1| \gtrsim 275~{\rm GeV}$, $M_2 \gtrsim 300~{\rm GeV}$, $\mu \gtrsim 460~{\rm GeV}$, $m_{\tilde{\mu}_L} \gtrsim 310~{\rm GeV}$, and $m_{\tilde{\mu}_R} \gtrsim 350~{\rm GeV}$, where $M_1$ and $M_2$ denote gaugino masses, $\mu$ represents the Higgsino mass, and $m_{\tilde{\mu}_L}$ and $m_{\tilde{\mu}_R}$ are the mass of Smuons with $L$ and $R$ denoting their dominant chiral component. This observation has significant impacts on dark matter (DM) physics, e.g., the popular $Z$- and Higgs-funnel regions have been excluded, and the Bino-dominated neutralino DM has to co-annihilate with the Wino-dominated electroweakinos (in most cases) and/or Smuons (in few cases) to obtain the correct density. It is also inferred that these conclusions should be applicable to the Minimal Supersymmetric Standard Model since the underlying physics for the bounds are the same.

\textbf{Keywords:} supersymmetric; muon anomalous magnetic moment; LHC

\textbf{PACS:} 04.65.+e 11.30.Pb 12.60.Jv 13.40.Em
}

\begin{document}
    \maketitle
    \flushbottom
\newpage
\section{Introduction}
The E989 experiment at Fermilab(FNAL) has released its first measurement of the muon anomalous magnetic moment $a_\mu \equiv (g-2)_\mu/2$ with 460 ppb precision~\cite{Abi:2021gix}, which corroborates the previous E821 result at the Brookhaven National Laboratory (BNL)~\cite{Bennett:2006fi} and increases the persistent tension between the experimental data and the Standard Model (SM) prediction. The averaged result after combining the BNL and FNAL data reads:

\vspace*{-0.4cm}

\begin{equation}
a_\mu^{\rm Exp} = 116592061(41)  \times 10^{-11},  \label{amu-exp}
\end{equation}
and it corresponds to a 4.2 $\sigma$ discrepancy from the current consensus of the SM prediction~\cite{Aoyama:2020ynm,Aoyama:2012wk,Aoyama:2019ryr,Czarnecki:2002nt,Gnendiger:2013pva,Davier:2017zfy,Keshavarzi:2018mgv,Colangelo:2018mtw,Hoferichter:2019gzf,Davier:2019can,	Keshavarzi:2019abf,Kurz:2014wya,Melnikov:2003xd,Masjuan:2017tvw,Colangelo:2017fiz,Hoferichter:2018kwz,Gerardin:2019vio,Bijnens:2019ghy,Colangelo:2019uex,Blum:2019ugy,Colangelo:2014qya}:

\vspace*{-0.8cm}

\begin{align}
a_\mu^{\rm SM} &= 116 591 810(43)  \times 10^{-11},\\
\Delta a_\mu &= a_\mu^{\rm Exp} - a_\mu^{\rm SM} = (251 \pm 59) \times 10^{-11}.  \label{delta-amu}
\end{align}
This excess is most likely to be further substantiated by a more thorough analysis at FNAL and J-PARC~\cite{Abe:2019thb} in the future. Although there are doubts that the muon anomaly may arise from uncertainties in experimental analysis and theoretical calculations~\cite{Cowan:2021sdy}, it is generally believed that the anomaly originates from new physics beyond the SM (BSM)~\cite{Athron:2021iuf}.  This speculation has motivated many BSM models to explain the anomaly, e.g., two Higgs doublet models, leptoquark models, vector-like lepton models, and so on (for a recent comprehensive study about this subject, see e.g.,~\cite{Athron:2021iuf}). Supersymmetry (SUSY), however, is one of the particular interests due to its elegant structure and natural solutions to many puzzles in the SM, such as the hierarchy problem, the unification of different forces, and the dark matter (DM) mystery~\cite{Fayet:1976cr,Haber:1984rc,Gunion:1984yn,Djouadi:2005gj,Martin:1997ns,Jungman:1995df}. In fact, many studies on the anomaly in low-energy supersymmetric models have demonstrated that the source of the significant deviation can be totally or partially attributed to the loop diagram corrections of supersymmetric particles (sparticles), i.e., smuon-neutralino loops and sneutrino-chargino loops~\cite{Moroi:1995yh,Hollik:1997vb,Czarnecki:2001pv,Stockinger:2006zn,Domingo:2008bb,Cao:2011sn,Athron:2015rva,Padley:2015uma,Kang:2016iok,Zhu:2016ncq,Okada:2016wlm,Yanagida:2017dao,Du:2017str,Ning:2017dng,
Hagiwara:2017lse,Choudhury:2017fuu,Cox:2018qyi,Tran:2018kxv,Wang:2018vxp,Yang:2018guw,Cao:2019evo,Pyarelal:2019zth,Liu:2020nsm,
Cao:2021lmj,Ke:2021kgy,Lamborn:2021snt,Li:2021xmw,Nakai:2021mha,Li:2021koa,Kim:2021suj,Li:2021pnt,Altmannshofer:2021hfu,Baer:2021aax,
Chakraborti:2021bmv,Aboubrahim:2021xfi,Athron:2021iuf,Iwamoto:2021aaf,Chakraborti:2021dli,Cao:2021tuh,Yin:2021mls,Zhang:2021gun,Ibe:2021cvf,Han:2021ify,
Wang:2021bcx,Chakraborti:2021mbr,Aboubrahim:2021myl,Ali:2021kxa,Wang:2021lwi,Chakraborti:2020vjp,Baum:2021qzx,Li:2021bbf,Forster:2021vyz,VanBeekveld:2021tgn,Zheng:2021wnu,Jeong:2021qey,
Martin:2001st,Endo:2021zal,Chakraborti:2022vds,Gomez:2022qrb,Chakraborti:2021kkr,Cao:2022chy}.

In the Minimal Supersymmetric Standard Model~(MSSM)~\cite{Haber:1984rc,Gunion:1984yn,Djouadi:2005gj}, the lightest neutralino $\tilde{\chi}_1^0$ is usually taken as the lightest supersymmetric particle~(LSP) and can behave like a weakly-interacting-massive-particle~(WIMP) DM candidate under the assumption of R-parity conservation~\cite{Farrar:1978xj,Jungman:1995df}. Specifically, the major component of $\tilde{\chi}_1^0$ should be Bino field since a Wino-like or Higgsino-like $\tilde{\chi}_1^0$ must be around $1~{\rm TeV}$ in mass to fully account for the correct DM relic density~\cite{Jungman:1995df}, and consequently, the other sparticles are heavy to make the theory incapable of explaining the anomaly. Recent researches have discussed the MSSM explanation of the anomaly within 2$\sigma$ uncertainty by keeping the theory consistent with the measurement of the DM relic abundance, the negative results from DM direct detection~(DD) experiments, and searches for electroweakinos  at the Large Hadron Collider~(LHC)~\cite{Chakraborti:2020vjp,Chakraborti:2021mbr,Chakraborti:2021squ,Chakraborti:2021dli}. Three scenarios classified by DM annihilation mechanisms were comprehensively analyzed~\cite{Chakraborti:2020vjp,Chakraborti:2021dli}. It was found that, under the assumption that $\tilde{\chi}_1^0$ makes up the full DM content of the universe, the improved $(g-2)_\mu$ data would bring an upper limit of roughly 600 GeV on the LSP and next-to-LSP (NLSP) mass, which set a clear target for future electroweakino searches at high-luminosity LHC and high-energy $e^+e^-$ colliders. This conclusion also applies to the Higgsino- and Wino-dominated LSP cases if the measured DM relic density is regarded as an upper bound~\cite{Chakraborti:2021kkr}.

Even though the MSSM can readily explain the g-2 anomaly, the combined constraints from the DM and SUSY search experiments would require massive Higgsinos~\cite{Chakraborti:2021dli}, leading up to a fine-tuning in predicting Z boson mass~\cite{Baer:2012uy}. This fact motivates us to study the Next-to-Minimal Supersymmetric Standard Model with a $\mathbb{Z}_3$-symmetry ($\mathbb{Z}_3$-NMSSM), which is another economical realization of SUSY~\cite{Ellwanger:2009dp,Maniatis:2009re}, in explaining the anomaly. This model extends the MSSM with a singlet superfield $\hat{S}$, and consequently it can dynamically generate the $\mu$-parameter of the MSSM, significantly enhance the SM-like Higgs boson mass, and predict much richer phenomenology than the MSSM (see, for example, Refs.~\cite{Cao:2012fz,Cao:2018rix}). One remarkable improvement of this study over the previous ones in~\cite{Moroi:1995yh,Hollik:1997vb,Czarnecki:2001pv,Stockinger:2006zn,Domingo:2008bb,Cao:2011sn,Athron:2015rva,Padley:2015uma,Kang:2016iok,Zhu:2016ncq,Okada:2016wlm,Yanagida:2017dao,Du:2017str,Ning:2017dng,
Hagiwara:2017lse,Choudhury:2017fuu,Cox:2018qyi,Tran:2018kxv,Wang:2018vxp,Yang:2018guw,Cao:2019evo,Liu:2020nsm,
Cao:2021lmj,Ke:2021kgy,Lamborn:2021snt,Li:2021xmw,Nakai:2021mha,Li:2021koa,Kim:2021suj,Li:2021pnt,Altmannshofer:2021hfu,Baer:2021aax,
Chakraborti:2021bmv,Aboubrahim:2021xfi,Iwamoto:2021aaf,Chakraborti:2021dli,Cao:2021tuh,Yin:2021mls,Zhang:2021gun,Ibe:2021cvf,Han:2021ify,
Wang:2021bcx,Chakraborti:2021mbr,Aboubrahim:2021myl,Ali:2021kxa,Wang:2021lwi,Chakraborti:2020vjp,Baum:2021qzx,Li:2021bbf,Forster:2021vyz,VanBeekveld:2021tgn,Zheng:2021wnu,Jeong:2021qey,
Martin:2001st,Endo:2021zal,Chakraborti:2022vds,Gomez:2022qrb,Chakraborti:2021kkr} is that more SUSY searches at the LHC, such as ATLAS-2106-01676~\cite{ATLAS:2021moa}, CMS-SUS-16-039~\cite{CMS:2017moi}, CMS-SUS-17-004~\cite{CMS:2018szt}, and CMS-SUS-21-001~\cite{CMS:2020bfa},  are included to limit theory parameter space. As a result, lower bounds on sparticle mass spectra are obtained, i.e.,  $|M_1| \gtrsim 275~{\rm GeV}$, $M_2 \gtrsim 300~{\rm GeV}$,  $\mu \gtrsim 460~{\rm GeV}$, $m_{\tilde{\mu}_L} \gtrsim 310~{\rm GeV}$, and $m_{\tilde{\mu}_R} \gtrsim 350~{\rm GeV}$, where $M_1$ and $M_2$ denote gaugino masses, $\mu$ represents the Higgsino mass, and $m_{\tilde{\mu}_L}$ and $m_{\tilde{\mu}_R}$ are the mass of Smuons with $L$ and $R$ denoting their dominant chiral component. These bounds are far beyond the reach of the LEP experiments in searching for SUSY and have not been emphasized before. They have significant impacts on DM physics, e.g., the popular $Z$- and Higgs-funnel regions are excluded, and the neutralino DM obtained the correct density mainly by co-annihilating with the Wino-dominated electroweakinos. In addition, it is inferred that these conclusions can also be applied to the MSSM since the underlying physics for the bounds are the same.

This work investigates the impact of the latest $(g-2)_{\mu}$ measurement and the LHC searches for SUSY on the $\mathbb{Z}_3$-NMSSM. It is organized as follows. In Section~\ref{theory-section}, we briefly introduce the SUSY contribution to the moment and the latest LHC probes of SUSY. In Section~\ref{numerical study}, we state the research strategy and analyze the numerical results. Lastly, we summarize the results of this study in Section~\ref{conclusion-section}.

\vspace{-0.3cm}

\section{\label{theory-section}Theoretical preliminaries}

\vspace{-0.2cm}

\subsection{\label{z3} $\mathbb{Z}_3$-NMSSM}

Compared with the MSSM, the $\mathbb{Z}_3$-NMSSM introduces a new gauge-singlet Higgs superfield $\hat{S}$. Its superpotential is given by~\cite{Maniatis:2009re,Ellwanger:2009dp}:
\begin{align}
\label{eq:superpotential}
W_\mathrm{\mathbb{Z}_{3}-NMSSM} = \lambda \hat{S} \hat{H_u} \cdot \hat{H_d} + \frac{1}{3} \kappa \hat{S}^3 + W_\mathrm{Yukawa},
\end{align}
where $\hat{H}_u$ and $\hat{H}_d$ represent the up- and down-type doublet Higgs superfield, respectively, $\lambda$ and $\kappa$ are dimensionless Yukawa parameters, and $W_\mathrm{Yukawa}$ denotes the Yukawa couplings that are the same as those in the MSSM. All the terms in $W_\mathrm{\mathbb{Z}_3-NMSSM}$ follow the $\mathbb{Z}_3$ symmetry. The corresponding soft-breaking terms for Eq.(\ref{eq:superpotential}) are:
\begin{align}\label{soft_term}
V_{\rm {\mathbb{Z}_{3}-NMSSM}}^{\rm soft}&= m_{H_u}^2|H_u|^2 + m_{H_d}^2|H_d|^2
+ m_S^2|S|^2 \nonumber \\ 
&+( \lambda A_{\lambda} SH_u\cdot H_d
+\frac{1}{3}\kappa A_{\kappa} S^3 + h.c.). 
\end{align}
In practice, the soft-breaking mass parameters $m_{H_u}^2,m_{H_d}^2$, and $m_S^2$ can be fixed by solving the electroweak symmetry breaking~(EWSB) equations, where the vacuum expectation values of the Higgs scalar fields are taken as $\left\langle H_u^0 \right\rangle = v_u/\sqrt{2}$, $\left\langle H_d^0 \right\rangle = v_d/\sqrt{2}$ and $\left\langle S \right\rangle = v_s/\sqrt{2}$ with $v = \sqrt{v_u^2+v_d^2}\simeq 246~\mathrm{GeV}$, $\tan{\beta} \equiv v_u/v_d$ and the effective $\mu$ parameter is generated by $\mu=\lambda v_s$. Consequently, the Higgs sector of $\mathbb{Z}_3$-NMSSM is described by the following six parameters~\cite{Ellwanger:2009dp}:
\begin{align}
\lambda,~ \kappa,~ A_\lambda,~ A_\kappa,~ \mu,~ \tan\beta.\nonumber
\label{eq:six}
\end{align}
The mixings of the fields \emph{\textbf{$H_u^0$, $H_d^0$}} and $S$ result in five mass eigenstates, including three CP-even Higgs $h_i$ ($i=1,2,3$ with $m_{h_1} < m_{h_2} < m_{h_3}$) and two CP-odd Higgs $A_j$ ($j=1,2$ with $m_{A_1} < m_{A_2}$). In this work, the lightest CP-even Higgs $h_1$ is treated as the SM-like Higgs boson since the Bayesian evidence of the $h_1$-scenario is much larger than that of the $h_2$-scenario after considering experimental constraints~\cite{Cao:2018iyk}.

The mixings between the $U(1)_{Y}$ and $SU(2)_{L}$ gaugino fields (Bino $\tilde{B}$ and Wino $\tilde{W}$), Higgsino fields ($\tilde{H_u}$ and $\tilde{H_d}$), and Singlino field $\tilde{S}$ form into five neutralinos $\tilde{\chi}^0_i$ ($i=1,2,...5$ and in an ascending mass order) and two charginos $\tilde{\chi}^\pm_j$ ($j=1,2$ with $m_{\tilde{\chi}_1^{\pm}}<m_{\tilde{\chi}_2^{\pm}}$). Their masses and mixings are determined by the soft-breaking gaugino masses $M_1$ and $M_2$, the Higgsino mass $\mu$, $\lambda$, $\kappa$, and $\tan\beta$.

\subsection{\label{DMRD}Muon $g-2$ in the $\mathbb{Z}_3$-NMSSM }

The SUSY contribution $a_{\mu}^{\rm{SUSY}}$ in the $\mathbb{Z}_3$-NMSSM comes from $\tilde{\mu}-\tilde{\chi}^0$ loops and $\tilde{\nu}_{\mu}-\tilde{\chi}^{\pm}$ loops~\cite{Domingo:2008bb,Martin:2001st}. The expression of the one-loop contribution to $a^{\rm SUSY}_{\mu}$ in the $\mathbb{Z}_3$-NMSSM is similar to that in the MSSM and given by~\cite{Domingo:2008bb}:
\begin{small}\begin{equation}\begin{split}
	&a_{\mu}^{\rm SUSY} = a_{\mu}^{\tilde{\chi}^0 \tilde{\mu}} + a_{\mu}^{\tilde{\chi}^{\pm} \tilde{\nu}},\\
	a_{\mu}^{\tilde{\chi}^0 \tilde{\mu}} &= \frac{m_{\mu}}{16 \pi^2}\sum_{i,l}\left\{
	-\frac{m_{\mu}}{12 m_{\tilde{\mu}_l}^2} \left( |n_{il}^{\rm L}|^2 + |n_{il}^{\rm R}|^2 \right) F_1^{\rm N}(x_{il}) + \frac{m_{\tilde{\chi}_i^0}}{3 m_{\tilde{\mu}_l}^2} {\rm Re}(n_{il}^{\rm L} n_{il}^{\rm R}) F_2^{\rm N}(x_{il})
	\right\}, \\
	a_{\mu}^{\tilde{\chi}^\pm \tilde{\nu}} &= \frac{m_{\mu}}{16 \pi^2}\sum_{k}\left\{
	\frac{m_{\mu}}{12 m_{\tilde{\nu}_{\mu}}^2} \left( |c_{k}^{\rm L}|^2 + |c_{k}^{\rm R}|^2 \right) F_1^{\rm C}(x_{k}) + \frac{2 m_{\tilde{\chi}_k^\pm}}{3 m_{\tilde{\nu}_{\mu}}^2} {\rm Re}(c_{k}^{\rm L}c_{k}^{ \rm R}) F_2^{\rm C}(x_{k})
	\right\}, \label{amuon}
	\end{split}
	\end{equation}\end{small}
where $i=1,\cdots,5$, $j=1,2$, and $l=1,2$ denote the neutralino, chargino and smuon index, respectively.
\begin{equation}
\begin{split}
n_{il}^{\rm L} 	= \frac{1}{\sqrt{2}}\left( g_2 N_{i2} + g_1 N_{i1} \right)X^*_{l1} -y_{\mu} N_{i3}X^*_{l2}, \quad
&n_{il}^{\rm R} = \sqrt{2} g_1 N_{i1} X_{l2} + y_{\mu} N_{i3} X_{l1},\\
c_{k}^{\rm L}  	= -g_2 V^{\rm c}_{k1}, \quad
&c_{k}^{\rm R} 	= y_{\mu} U^{\rm c}_{k2}, \\
\end{split}
\end{equation}
where $N$ is the neutralino mass rotation matrix, $X$ the smuon mass rotation matrix, and $U^{\rm c}$ and $V^{\rm c}$ the chargino mass rotation matrices defined by ${U^{\rm c}}^* M_{C} {V^{\rm c}}^\dag = m_{\tilde{\chi}^\pm}^{\rm diag}$. $F(x)$s are the loop functions of the kinematic variables defined as $x_{il} \equiv m_{\tilde{\chi}_i^0}^2 / m_{\tilde{\mu}_l}^2$ and $x_{k} \equiv m_{\tilde{\chi}_k^\pm}^2 / m_{\tilde{\nu}_{\mu}}^2$, and take the form:
\begin{align}
F^N_1(x) & = \frac{2}{(1-x)^4}\left[ 1-6x+3x^2+2x^3-6x^2\ln x\right] \\
F^N_2(x) & = \frac{3}{(1-x)^3}\left[ 1-x^2+2x\ln x\right] \\
F^C_1(x) & = \frac{2}{(1-x)^4}\left[ 2+ 3x - 6x^2 + x^3 +6x\ln x\right] \\
F^C_2(x) & = -\frac{3}{2(1-x)^3}\left[ 3-4x+x^2 +2\ln x\right].
\end{align}
They satisfy $F^N_1(1) = F^N_2(1) = F^C_1(1) = F^C_2(1) = 1$ for the mass-degenerate sparticle case.

It is helpful to understand the features of $a_\mu^{\rm SUSY}$  with the mass insertion calculation method~\cite{Moroi:1995yh}. In the lowest order approximation, the contributions of $a_\mu^{\rm SUSY}$ can be classified into four types: "WHL", "BHL", "BHR" and "BLR" where W, B, H, L and R denote the Wino, Bino, Higgsino, left-handed Smuon (or Sneutrino) and right-handed Smuon field, respectively. They can be expressed as~\cite{Athron:2015rva,Moroi:1995yh,Endo:2021zal}:
\begin{eqnarray}
a_{\mu, \rm WHL}^{\rm SUSY}
&=&\frac{\alpha_2}{8 \pi} \frac{m_{\mu}^2 M_2 \mu \tan \beta}{M_{\tilde{\nu}_\mu}^4} \left \{ 2 f_C\left(\frac{M_2^2}{M_{\tilde{\nu}_{\mu}}^2}, \frac{\mu^2}{M_{\tilde{\nu}_{\mu}}^2} \right) - \frac{M_{\tilde{\nu}_\mu}^4}{M_{\tilde{\mu}_L}^4} f_N\left(\frac{M_2^2}{M_{\tilde{\mu}_L}^2}, \frac{\mu^2}{M_{\tilde{\mu}_L}^2} \right) \right \}\,, \quad \quad
\label{eq:WHL} \\
a_{\mu, \rm BHL}^{\rm SUSY}
&=& \frac{\alpha_Y}{8 \pi} \frac{m_\mu^2 M_1 \mu  \tan \beta}{M_{\tilde{\mu}_L}^4} f_N\left(\frac{M_1^2}{M_{\tilde{\mu}_L}^2}, \frac{\mu^2}{M_{\tilde{\mu}_L}^2} \right)\,,
\label{eq:BHL} \\
a_{\mu, \rm BHR}^{\rm SUSY}
&=& - \frac{\alpha_Y}{4\pi} \frac{m_{\mu}^2 M_1 \mu \tan \beta}{M_{\tilde{\mu}_R}^4} f_N\left(\frac{M_1^2}{M_{\tilde{\mu}_R}^2}, \frac{\mu^2}{M_{\tilde{\mu}_R}^2} \right)\,,
\label{eq:BHR} \\
a_{\mu \rm BLR}^{\rm SUSY}
&=& \frac{\alpha_Y}{4\pi} \frac{m_{\mu}^2  M_1 \mu \tan \beta}{M_1^4}
f_N\left(\frac{M_{\tilde{\mu}_L}^2}{M_1^2}, \frac{M_{\tilde{\mu}_R}^2}{M_1^2} \right)\,,
\label{eq:BLR}
\end{eqnarray}
where $M_{\tilde{\mu}_L}$ and $M_{\tilde{\mu}_R}$ are the masses for left- and right-handed Smuon field, respectively.
The loop function $f_C$ and $f_N$ take the following forms:
\begin{eqnarray}
\label{eq:loop-aprox}
f_C(x,y)
&=&  \frac{5-3(x+y)+xy}{(x-1)^2(y-1)^2} - \frac{2\ln x}{(x-y)(x-1)^3}+\frac{2\ln y}{(x-y)(y-1)^3} \,,
\\
f_N(x,y)
&=&
\frac{-3+x+y+xy}{(x-1)^2(y-1)^2} + \frac{2x\ln x}{(x-y)(x-1)^3}-\frac{2y\ln y}{(x-y)(y-1)^3}.
\end{eqnarray}
The expressions of $a_{\mu,\ i}^{\rm SUSY}$ (i= WHL, BHL, BHR) involve a prefactor of the Higgsino mass $\mu$ as well as the loop functions which approach zero with the increase of $|\mu|$. Consequently, they depend on $\mu$ in a complex way. For several typical patterns of sparticle mass spectra with a positive $\mu$, it is found that the "WHL" contribution decreases monotonously as $\mu$ increases, while the magnitude of the "BHL" and "BHR" contributions increases when $\mu$ is significantly smaller than the slepton mass and decreases when $\mu$ is larger than the slepton mass. In addition, it should be noted that the "WHL" contribution is usually the dominant one when $M_{\tilde{\mu}_L}$ is not significantly larger than $M_{\tilde{\mu}_R}$\footnote{We are not interested in the excessively large $|\mu|$ case, where the "BLR" contribution is the dominant one, because it needs fine tunings of SUSY parameters in predicting $Z$-boson mass~\cite{Baer:2012uy}. }. It should also be noted that, since the Singlino field only appears in the "WHL", "BHL" and "BHR" loops by two more insertions at the lowest order, its induced contribution to $a_{\mu}^{\rm{SUSY}}$ is less prominent. Thus the predictions for $a_\mu^{\rm SUSY}$ in the NMSSM is almost the same as that in the MSSM. Even so, the two theories may still display different features in fitting experimental constraints due to their possibly distinct DM physics and sparticle signals at the LHC~\cite{Cao:2021tuh,Cao:2022chy}.

\subsection{\label{analyses}LHC Analyses}

Since the electroweakinos and sleptons involved in $a_\mu^{\rm SUSY}$ are not very heavy to explain the anomaly (see the results presented below), they can be copiously produced at the LHC, and thus are subjected to strong constraints from the  analyses of the experimental data at $\sqrt{s}=13~\rm{TeV}$. Given the complexity of their production processes and decay modes, many signal topologies should be studied. It was found that the following analyses are particularly important for this work:
\begin{itemize}
	\item \texttt{CMS-SUS-16-039 and CMS-SUS-17-004~\cite{CMS:2017moi,CMS:2018szt}}: Search for electroweakino productions in the pp collisions with two, three, or four leptons and missing transverse momentum ($\rm{E}_{\rm{T}}^{\rm{miss}}$) as the final states. Given the smallness of the production cross-sections, the analyses included all the possible final states and defined several categories by the number of leptons in the event, their flavors, and their charges to enhance the discovery potential. The results were interpreted in the context of simplified models for either Wino-like chargino-neutralino production or neutralino pair production in a gauge-mediated SUSY breaking (GMSB) scenario. An observed (expected) limit on $m_{\tilde{\chi}_1^{\pm}}$ in the chargino-neutralino production was about 650 (570) GeV for the WZ topology, 480 (455) GeV for the WH topology, and 535 (440) GeV for the mixed topology. Instead, the observed and expected limits on the neutralino mass in the GMSB scenario were 650–750 GeV and 550–750 GeV, respectively.
    \item \texttt{CMS-SUS-20-001~\cite{CMS:2020bfa}}: Search for $2~\rm{leptons} + \rm{jets} + \rm{E}_{\rm{T}}^{\rm{miss}}$ signal. Specifically, four scenarios were looked closely. The first one targeted strong sparticle productions with at least one on-shell $Z$ boson in the decay chain. Six disjoint categories were defined by the number of jets (i.e., SRA, SRB and SRC), which were reconstructed by requiring the distance parameter less than 0.4 and $p_{\rm{T}}^{j} \geq 35~\rm{GeV}$, and whether the presence of b-tagged jets. The second one also required the decay chain to contain an on-shell $Z$ boson, but it scrutinized the electroweakino production.  It defined the VZ category by the decay modes $Z Z \to (\ell \bar{\ell}) (q \bar{q})$ and $Z W \to (\ell \bar{\ell}) (q \bar{q}^\prime)$, and the HZ category by $Z h \to (\ell \bar{\ell}) (b \bar{b})$, where $h$ denoted the SM Higgs boson. The third one, referred to as the "edge" scenario, investigated the strong production with an off-shell $Z$ boson or a slepton in the decay chain. It required two or more jets, $p_{\rm{T}}^{\rm{miss}} > 150$ or $200~\rm{GeV}$, and $M_{\rm{T}_2}(\ell\ell) > 80 ~\rm{GeV}$ in its signal regions. The last one studied slepton pair production by examining the signal with two leptons, $p_{\rm{T}}^{\rm{miss}} > 100~\rm{GeV}$, no b-tagged jets, and moderate jet activity. This analysis excluded sparticles up to 1870 GeV in mass for Gluinos, 1800 GeV for light-flavor Squarks, 1600 GeV for bottom Squarks, 750 GeV and 800 GeV for Wino-dominated chargino and neutralino, respectively, and 700 GeV for the first two-generation Sleptons.

   \item \texttt{ATLAS-2106-01676~\cite{ATLAS:2021moa}}: Search for Higgsino- and Wino-dominated chargino-neutralino production, including the cases of compressed and non-compressed mass spectra. This analysis studied on-shell $WZ$, off-shell $WZ$, and $Wh$ scenarios, and required the final states to contain exactly three leptons, possible ISR jets, and $\rm{E}_{\rm{T}}^{\rm{miss}}$. For the Higgsino model, $\tilde{\chi}_2^0$ was excluded up to $210~\rm{ GeV}$ in mass for the off-shell W/Z case; while for the Wino model, the exclusion bound of $\tilde{\chi}_2^0$ was $640~\rm{GeV}$ and $300~\rm{GeV}$ for the on-shell and off-shell W/Z case, respectively.
    \item \texttt{ATLAS-1908-08215~\cite{ATLAS:2019lff}}: Search for chargino pair and slepton pair productions with two leptons
    and missing transverse momentum as their final state. This analysis considered the following three simplified models: $p p \to \tilde{\chi}_1^\pm \tilde{\chi}_1^\mp \to (W^\pm \tilde{\chi}_1^0) (W^\mp \tilde{\chi}_1^0)$, $p p \to \tilde{\chi}_1^\pm \tilde{\chi}_1^\mp \to (\tilde{\ell}^\ast \nu_\ell) (\bar{\nu}_\ell \tilde{\ell}), (\bar{\ell} \tilde{\nu}_\ell) (\tilde{\nu}^\ast_\ell \ell)$, and $p p \to \tilde{\ell}^\ast \tilde{\ell} \to (\bar{\ell} \tilde{\chi}_1^0) (\ell \tilde{\chi}_1^0)$.
    For a massless $\tilde{\chi}_1^0$, $\tilde{\chi}_1^\pm$ could be excluded up to 420 GeV and 1 TeV, respectively, and the slepton excluded up to 700 GeV, assuming that sleptons are mass-degenerated in flavor and chiral space.
    \item \texttt{ATLAS-1911-12606~\cite{ATLAS:2019lng}}: Concentrate on the case of compressed mass spectra and search for the electroweakino production with two leptons and missing transverse momentum as the final state. Four scenarios were used to interpret the analyses. The first one studied $\tilde{\chi}_1^{\pm}\tilde{\chi}_1^{\mp}$, $\tilde{\chi}_2^0\tilde{\chi}_1^{\pm}$ and $\tilde{\chi}_2^0\tilde{\chi}_1^0$ productions in the Higgsino model. The results were projected onto $\Delta m-\tilde{\chi}_2^0$ plane where $\Delta m \equiv m_{\tilde{\chi}_2^0} -  m_{\tilde{\chi}_1^0}$. It was found that the tightest bound on $m_{\tilde{\chi}_2^0}$ was $193~{\rm GeV}$ for $\Delta m \simeq 9.3~{\rm GeV}$. The second scenario was quite similar to the first one except for the $\tilde{\chi}_2^0 \tilde{\chi}_1^\pm$ production in the Wino/Bino model.  The optimum bound on $m_{\tilde{\chi}_2^0}$ was $240~{\rm GeV}$ when $\Delta m \simeq 7~{\rm GeV}$. The third one assumed that the electroweakino pair production proceeded via the vector-boson fusion (VBF) process and used the kinematic cuts on  $m_{\ell\ell}$ as the primary discriminator. Correspondingly, constraints on the $\Delta m-\tilde{\chi}_2^0$ plane were obtained for both the Higgsino and Wino/Bino models, which were significantly weaker than the previous results. The last one targeted the slepton pair production. It exploited the relationship between the lepton momenta and the missing transverse momentum through the transverse mass, $m_{T2}$, which exhibited a kinematic endpoint similar to that for $m_{\ell\ell}$ in the electroweakino decays. Light-flavor sleptons were found to be heavier than about 250 GeV for $\Delta m_{\tilde{\ell}} = 10~{\rm GeV}$, where $m_{\tilde{\ell}} \equiv m_{\tilde{\ell}} - m_{\tilde{\chi}_1^0}$.
\end{itemize}
Concerning these analyses, it should be noted that only the first one studied the data obtained with $36~{\rm fb}^{-1}$ integrated luminosity, and the others were based on $139~{\rm fb}^{-1}$ data.

\begin{table}[]
	\caption{Experimental analyses included in the package \texttt{SModelS-2.1.1}.}
	\label{tab:1}
	\vspace{0.1cm}
	\resizebox{1.0 \textwidth}{!}{
		\begin{tabular}{cccc}
			\hline\hline
			\texttt{Name} & \texttt{Scenario} &\texttt{Final State} &$\texttt{Luminosity} (\texttt{fb}^{\texttt{-1}})$ \\\hline
			\begin{tabular}[l]{@{}l@{}} CMS-SUS-17-010~\cite{CMS:2018xqw}\end{tabular}   &\begin{tabular}[c]{@{}c@{}}$\tilde{\chi}_1^{\pm}\tilde{\chi}_1^{\mp}\rightarrow W^{\pm}\tilde{\chi}_1^0 W^{\mp}\tilde{\chi}_1^0$\\$\tilde{\chi}_1^{\pm}\tilde{\chi}_1^{\mp}\rightarrow \nu\tilde{\ell} \ell\tilde{\nu}$ \\ \end{tabular}&2$ \ell$  + $E_{\rm T}^{\rm miss}$    & 35.9  \\ \\
			\begin{tabular}[l]{@{}l@{}} CMS-SUS-17-009~\cite{CMS:2018eqb}\end{tabular}   &$\tilde{\ell}\tilde{\ell}\rightarrow \ell\tilde{\chi}_1^0\ell\tilde{\chi}_1^0$ &2$\ell$ + $E_{\rm T}^{\rm miss}$    &  35.9               \\ \\
			\begin{tabular}[l]{@{}l@{}} CMS-SUS-17-004~\cite{CMS:2018szt}\end{tabular} &$\tilde{\chi}_{2}^0\tilde{\chi}_1^{\pm}\rightarrow Wh(Z)\tilde{\chi}_1^0\tilde{\chi}_1^0$ & n$ \ell$(n\textgreater{}=0) + nj(n\textgreater{}=0) + $ E_{\rm T}^{\rm miss}$   & 35.9               \\ \\
			\begin{tabular}[l]{@{}l@{}}CMS-SUS-16-045~\cite{CMS:2017bki}\end{tabular}          &$\tilde{\chi}_2^0\tilde{\chi}_1^{\pm}\rightarrow W^{\pm}\tilde{\chi}_1^0h\tilde{\chi}_1^0$& 1$ \ell$ 2b + $ E_{\rm T}^{\rm miss}$                           & 35.9               \\ \\
			\begin{tabular}[l]{@{}l@{}} CMS-SUSY-16-039~\cite{CMS:2017moi} \end{tabular}          &\begin{tabular}[c]{@{}c@{}c@{}c@{}c@{}} $\tilde{\chi}_2^0\tilde{\chi}_1^{\pm}\rightarrow \ell\tilde{\nu}\ell\tilde{\ell}$\\$\tilde{\chi}_2^0\tilde{\chi}_1^{\pm}\rightarrow\tilde{\tau}\nu\tilde{\ell}\ell$\\$\tilde{\chi}_2^0\tilde{\chi}_1^{\pm}\rightarrow\tilde{\tau}\nu\tilde{\tau}\tau$\\ $\tilde{\chi}_2^0\tilde{\chi}_1^{\pm}\rightarrow WZ\tilde{\chi}_1^0\tilde{\chi}_1^0$\\$\tilde{\chi}_2^0\tilde{\chi}_1^{\pm}\rightarrow WH\tilde{\chi}_1^0\tilde{\chi}_1^0$\end{tabular} & n$\ell(n\textgreater{}0)$($\tau$) + $E_{\rm T}^{\rm miss}$& 35.9\\ \\
			\begin{tabular}[l]{@{}l@{}}CMS-SUS-16-034~\cite{CMS:2017kxn}\end{tabular}&$\tilde{\chi}_2^0\tilde{\chi}_1^{\pm}\rightarrow W\tilde{\chi}_1^0Z(h)\tilde{\chi}_1^0$ & n$\ell$(n\textgreater{}=2) + nj(n\textgreater{}=1) $E_{\rm T}^{\rm miss}$       &               35.9               \\ \\
			\begin{tabular}[l]{@{}l@{}}ATLAS-1803-02762~\cite{ATLAS:2018ojr}\end{tabular} &\begin{tabular}[c]{@{}c@{}c@{}c@{}}$\tilde{\chi}_2^0\tilde{\chi}_1^{\pm}\rightarrow WZ\tilde{\chi}_1^0\tilde{\chi}_1^0$\\$\tilde{\chi}_2^0\tilde{\chi}_1^{\pm}\rightarrow \nu\tilde{\ell}l\tilde{\ell}$\\$\tilde{\chi}_1^{\pm}\tilde{\chi}_1^{\mp}\rightarrow \nu\tilde{\ell}\nu\tilde{\ell}$\\ $ \tilde{\ell}\tilde{\ell}\rightarrow \ell\tilde{\chi}_1^0\ell\tilde{\chi}_1^0$\end{tabular} & n$ \ell$ (n\textgreater{}=2) + $ E_{\rm T}^{\rm miss}$ & 36.1               \\ \\
			\begin{tabular}[l]{@{}l@{}}ATLAS-1812-09432~\cite{ATLAS:2018qmw}\end{tabular} &$\tilde{\chi}_2^0\tilde{\chi}_1^{\pm}\rightarrow Wh\tilde{\chi}_1^0\tilde{\chi}_1^0$ & n$ \ell$ (n\textgreater{}=0) + nj(n\textgreater{}=0) + nb(n\textgreater{}=0) + n$\gamma$(n\textgreater{}=0) + $E_{\rm T}^{\rm miss}$ & 36.1               \\ \\
			\begin{tabular}[l]{@{}l@{}}ATLAS-1806-02293~\cite{ATLAS:2018eui}\end{tabular} &$\tilde{\chi}_2^0\tilde{\chi}_1^{\pm}\rightarrow WZ\tilde{\chi}_1^0\tilde{\chi}_1^0$ &n$\ell$(n\textgreater{}=2) + nj(n\textgreater{}=0) + $ E_T^{miss}$ & 36.1               \\ \\
			\begin{tabular}[l]{@{}l@{}}ATLAS-1912-08479~\cite{ATLAS:2019wgx}\end{tabular}          &$\tilde{\chi}_2^0\tilde{\chi}_1^{\pm}\rightarrow W(\rightarrow l\nu)\tilde{\chi}_1^0Z(\rightarrow\ell\ell)\tilde{\chi}_1^0$& 3$\ell $ + $ E_{\rm T}^{\rm miss}$                           & 139               \\ \\
			\begin{tabular}[l]{@{}l@{}}ATLAS-1908-08215~\cite{ATLAS:2019lff}\end{tabular}   &\begin{tabular}[c]{@{}c@{}}$\tilde{\ell}\tilde{\ell}\rightarrow \ell\tilde{\chi}_1^0\ell\tilde{\chi}_1^0$\\$\tilde{\chi}_1^{\pm}\tilde{\chi}_1^{\mp}$ \\ \end{tabular} & 2$\ell$ + $ E_{\rm T}^{\rm miss}$ & 139               \\ \\
			\begin{tabular}[l]{@{}l@{}}ATLAS-1909-09226~\cite{Aad:2019vvf}\end{tabular}          & $\tilde{\chi}_{2}^0\tilde{\chi}_1^{\pm}\rightarrow Wh\tilde{\chi}_1^0\tilde{\chi}_1^0$                            & 1$ \ell$ + h($\bm \rightarrow$ bb) + $ E_{\rm T}^{\rm miss}$    & 139               \\ \hline\\
			
	\end{tabular}} % }
\end{table}

\begin{table}[]
	\caption{Experimental analyses used in this study. All the analyses have been implemented in~\texttt{CheckMATE-2.0.29}, and some of them were finished by us.}
	\label{tab:2}
	\vspace{0.1cm}
	\resizebox{1.0 \textwidth}{!}{
		\begin{tabular}{cccc}
			\hline\hline
			\texttt{Name} & \texttt{Scenario} &\texttt{Final State} &$\texttt{Luminosity} (\texttt{fb}^{\texttt{-1}})$ \\\hline
			\texttt{ATLAS-1909-09226}~\cite{Aad:2019vvf}                  & $\tilde{\chi}_{2}^0\tilde{\chi}_1^{\pm}\rightarrow Wh\tilde{\chi}_1^0\tilde{\chi}_1^0$ & $1\ell + h(h\rightarrow bb) + \text{E}_\text{T}^{\text{miss}}$                  & 139                  \\\\
			\multirow{3}{*}{\texttt{ATLAS-1911-12606}~\cite{ATLAS:2019lng}} &$\tilde{\ell}\tilde{\ell}j\rightarrow \ell\tilde{\chi}_1^0 \ell\tilde{\chi}_1^0j$& \multirow{3}{*}{$2\ell + nj(n\textgreater{}=0) + \text{E}_\text{T}^{\text{miss}}$} & \multirow{3}{*}{139} \\
			&$(\text{Wino})\tilde{\chi}_2^0\tilde{\chi}_1^{\pm}j\rightarrow W^{\star}Z^{\star}\tilde{\chi}_1^0\tilde{\chi}_1^{0}j$&                    &                    \\
			&$(\text{Higgsino})\tilde{\chi}_2^0\tilde{\chi}_1^{\pm}j$ + $\tilde{\chi}_1^{+}\tilde{\chi}_1^{-}j$ + $\tilde{\chi}_2^0\tilde{\chi}_1^{0}j$&                    &    \\\\
			\texttt{CMS-SUS-20-001}~\cite{CMS:2020bfa}& $\tilde{\chi}_{2}^0\tilde{\chi}_1^{\pm}\rightarrow WZ\tilde{\chi}_1^0\tilde{\chi}_1^0$                            & $2\ell + nj(n\textgreater{}0) + \text{E}_\text{T}^{\text{miss}}$    & 137                  \\\\
			\multirow{2}{*}{\texttt{ATLAS-1908-08215}~\cite{ATLAS:2019lff}} &$\tilde{\ell}\tilde{\ell}\rightarrow \ell\tilde{\chi}_1^0\ell\tilde{\chi}_1^0$& \multirow{2}{*}{$2\ell + \text{E}_{\text{T}}^{\text{miss}}$} & \multirow{2}{*}{139} \\
			&$\tilde{\chi}_1^{\pm}\tilde{\chi}_1^{\mp}(\tilde{\chi}_1^{\pm}\rightarrow \tilde{\ell}\nu/\tilde{\nu}\ell)$&                  &                    \\\\
			\texttt{ATLAS-2106-01676}~\cite{ATLAS:2021moa}                &$\tilde{\chi}_2^0\tilde{\chi}_1^{\pm}\rightarrow W^{(*)}Z^{(*)}\tilde{\chi}_1^0\tilde{\chi}_1^0$,$\tilde{\chi}_2^0\tilde{\chi}_1^{\pm}\rightarrow Wh\tilde{\chi}_1^0\tilde{\chi}_1^0$&$3\ell + \text{E}_\text{T}^{\text{miss}}$ & 139 \\\\
			\multirow{3}{*}{\texttt{ATLAS-1803-02762}~\cite{ATLAS:2018ojr}}                &$\tilde{\chi}_2^0\tilde{\chi}_1^{\pm}\rightarrow WZ\tilde{\chi}_1^0\tilde{\chi}_1^0$,$\nu\tilde{\ell}l\tilde{\ell}$&\multirow{3}{*}{n$ \ell$ (n\textgreater{}=2) + $E_{\rm T}^{\rm miss}$}&\multirow{3}{*}{36.1} \\&$\tilde{\chi}_1^{\pm}\tilde{\chi}_1^{\mp}\rightarrow \nu\tilde{\ell}\nu\tilde{\ell}$&\\& $ \tilde{\ell}\tilde{\ell}\rightarrow\ell\tilde{\chi}_1^0\ell\tilde{\chi}_1^0$\\\\
			\multirow{5}{*}{\texttt{ATLAS-1802-03158}~\cite{ATLAS:2018nud}}                &$\tilde{g}\tilde{g}\rightarrow 2q\tilde{\chi}_1^0 2q\tilde{\chi}_1^0(\rightarrow \gamma \tilde{G})$&\multirow{5}{*}{n$ \gamma$ (n\textgreater{}=1) + nj(n\textgreater{}=0) + $E_{\rm T}^{\rm miss}$}&\multirow{5}{*}{36.1} \\&$\tilde{g}\tilde{g}\rightarrow 2q\tilde{\chi}_1^0(\rightarrow \gamma \tilde{G}) 2q\tilde{\chi}_1^0(\rightarrow Z \tilde{G})$\\&$\tilde{q}\tilde{q}\rightarrow q\tilde{\chi}_1^0(\rightarrow \gamma \tilde{G}) q\tilde{\chi}_1^0(\rightarrow \gamma \tilde{G})$&\\&$\tilde{\chi}_2^0\tilde{\chi}_1^{\pm}\rightarrow Z/h\tilde{\chi}_1^0 W\tilde{\chi}_1^0$\\& $\tilde{\chi}_1^{\pm}\tilde{\chi}_1^{\pm}\rightarrow W\tilde{\chi}_1^0 W\tilde{\chi}_1^0$  \\\\
			\multirow{3}{*}{\texttt{ATLAS-1712-08119}~\cite{ATLAS:2017vat}} &$\tilde{\ell}\tilde{\ell}\rightarrow \ell\tilde{\chi}_1^0\ell\tilde{\chi}_1^0$& \multirow{3}{*}{$2\ell + nj(n\textgreater{}=0) + \text{E}_\text{T}^{\text{miss}}$} & \multirow{3}{*}{36.1} \\
			&$(\text{Wino})\tilde{\chi}_2^0\tilde{\chi}_1^{\pm}\rightarrow WZ\tilde{\chi}_1^0\tilde{\chi}_1^0$&                    &                    \\
			&$(\text{Higgsino})\tilde{\chi}_2^0\tilde{\chi}_1^{\pm}$ + $\tilde{\chi}_1^{+}\tilde{\chi}_1^{-}$ + $\tilde{\chi}_2^0\tilde{\chi}_1^0$&                    &                    \\\\
			\multirow{2}{*}{\texttt{CMS-SUS-17-004}~\cite{CMS:2018szt}} &$\tilde{\chi}_2^0\tilde{\chi}_1^{\pm}\rightarrow WZ\tilde{\chi}_1^0\tilde{\chi}_1^0$,$ WH\tilde{\chi}_1^0\tilde{\chi}_1^0$& \multirow{2}{*}{$n\ell(n\textgreater{}0) + \text{E}_\text{T}^{\text{miss}}$}&\multirow{2}{*}{35.9} \\
			&$\tilde{\chi}_1^0\tilde{\chi}_1^{0}\rightarrow ZZ\tilde{G}\tilde{G}$,$HZ\tilde{G}\tilde{G}$,$HH\tilde{G}\tilde{G}$& &\\\\
			\multirow{3}{*}{\texttt{CMS-SUS-16-048}~\cite{CMS:2018kag}} &$\tilde{t}\tilde{t}\rightarrow b\tilde{\chi}_1^{\pm}b\tilde{\chi}_1^{\pm}$& \multirow{3}{*}{$n\ell(n\textgreater{}=0) + nb(n\textgreater{}=0) + nj(n\textgreater{}=0) + \text{E}_\text{T}^{\text{miss}}$} & \multirow{3}{*}{35.9} \\
			&$\tilde{\chi}_2^0\tilde{\chi}_1^{\pm}\rightarrow W^{*}Z^{*}\tilde{\chi}_1^0\tilde{\chi}_1^0$&                    &                    \\
			&$(\text{Higgsino})\tilde{\chi}_2^0\tilde{\chi}_1^{\pm}/\tilde{\chi}_1^0$&                    &    \\\\
			\multirow{3}{*}{\texttt{CMS-SUS-PAS-16-025}~\cite{CMS:2016zvj}} &$\tilde{t}\tilde{t}\rightarrow b\tilde{\chi}_1^{\pm}b\tilde{\chi}_1^{\pm}$& \multirow{3}{*}{$n\ell(n\textgreater{}=0) + nb(n\textgreater{}=0) + nj(n\textgreater{}=0) + \text{E}_\text{T}^{\text{miss}}$} & \multirow{3}{*}{12.9} \\
			&$\tilde{\chi}_2^0\tilde{\chi}_1^{\pm}\rightarrow W^{*}Z^{*}\tilde{\chi}_1^0\tilde{\chi}_1^0$&                    &                    \\\\
			&$(\text{Higgsino})\tilde{\chi}_2^0\tilde{\chi}_1^{\pm}/\tilde{\chi}_1^0$&                    &    \\\\
			\multirow{2}{*}{\texttt{ATLAS-CONF-2016-096}~\cite{ATLAS:2016uwq}} &$\tilde{\chi}_1^{\pm}\tilde{\chi}_1^{\mp}(\tilde{\chi}_1^{\pm}\rightarrow \tilde{\ell}\nu/\tilde{\nu}\ell)$& \multirow{2}{*}{$n\ell(n\textgreater{}=2) + \text{E}_{\text{T}}^{\text{miss}}$} & \multirow{2}{*}{13.3} \\\\
			&$\tilde{\chi}_1^{\pm}\tilde{\chi}_2^{0}(\tilde{\chi}_1^{\pm}\rightarrow \tilde{\ell}\nu/\tilde{\nu}\ell, \tilde{\chi}_2^{0}\rightarrow \tilde{\ell}\ell/\tilde{\nu}\nu)$& &\\  \hline
			                                     		
	\end{tabular}} % }
\end{table}

\vspace{-0.4cm}

\section{\label{numerical study} Explaining $\Delta a_\mu$ in $\mathbb{Z}_3$-NMSSM}

\vspace{-0.2cm}

\subsection{\label{scan} Research strategies}

In order to find out the features of the $\mathbb{Z}_3$-NMSSM in explaining the anomaly, a sophisticated scan was performed by the \texttt{MultiNest} algorithm~\cite{Feroz:2008xx} with $n_{\rm live}=8000$ in this study\footnote{The parameter $n_{\rm live}$ in the \texttt{MultiNest} algorithm controls the number of active points sampled from the prior distribution in each iteration of the scan.}. The explored parameter space was given by:
\begin{eqnarray}\begin{split}
\label{amuNMSSM-scan}
&  0 < \lambda \leq 0.7, ~~|\kappa| \leq 0.7, ~~1 \leq \tan \beta \leq 60,~~100 ~{\rm GeV} \leq \mu \leq 1~{\rm TeV}, 
\\
& |A_\kappa| \leq 1 ~{\rm TeV}, ~~|A_t| \leq 5~{\rm TeV},~~10~{\rm GeV} \leq A_\lambda \leq 5~{\rm TeV},~~|M_1| \leq 1.5~{\rm TeV},    \\
& 100~\rm{GeV} \leq M_2 \leq 1.5~\rm{TeV}, 100~\rm{GeV} \leq \tilde{M}_{\tilde{\mu}_L} \leq 1~\rm{TeV}, 100~\rm{GeV} \leq \tilde{M}_{\tilde{\mu}_R} \leq 1~\rm{TeV}, 
\end{split}\end{eqnarray}
where $\tilde{M}_{\tilde{\mu}_L}$ and $\tilde{M}_{\tilde{\mu}_R}$ denotes the soft-breaking mass of the left- and right-handed Smuon, respectively.
The Gluino mass was fixed at $M_3=3\tev$. The other dimensional parameters that were not crucial to this work were set to 2 TeV, including $A_\mu$ and the soft-breaking masses and soft trilinear coefficients for all squarks and the first- and third-generation sleptons. All the input parameters were defined at the renormalization scale $Q = 1~{\rm TeV}$ and followed flat prior distributions.

In the numerical calculation, the $\mathbb{Z}_3$-NMSSM model file was constructed by package \texttt{SARAH-4.14.3}~\cite{Staub:2008uz, Staub:2012pb, Staub:2013tta, Staub:2015kfa}. Particle mass spectra and low-energy observables, such as $a_\mu^{\rm SUSY}$ and B-physics observables, were calculated by programs \texttt{SPheno- 4.0.4}~\cite{Porod:2003um, Porod:2011nf} and \texttt{FlavorKit}~\cite{Porod:2014xia}. The DM abundance and direct/indirect detection cross-sections were obtained by package~\texttt{micrOMEGAs-5.0.4}~\cite{Belanger:2001fz, Belanger:2005kh, Belanger:2006is, Belanger:2010pz, Belanger:2013oya, Barducci:2016pcb}. The likelihood function that guided the scan process was dominated by the Gaussian distribution of $a_\mu^{\rm SUSY}$, which was expressed as:
\begin{equation}
\label{likelihood}
\mathcal{L}_{a_\mu^{\rm SUSY}}=\exp\left[-\frac{1}{2} \left( \frac{a_{\mu}^{\rm SUSY}- 2.51\times 10^{-9}}{5.9\times 10^{-10} }\right)^2\right],
\end{equation}
where the central value and error were taken from the combined result in Eq.(\ref{delta-amu}). Concerning experimental constraints, the likelihood function was set to 1 if the corresponding experimental limit was satisfied, otherwise, it took $\rm exp[-100]$ as a penalty. These constraints included:
\begin{itemize}
	\item {\bf DM relic density.}  Samples were required to predict the correct DM relic density with $0.096 \leq \Omega h^2 \leq 0.144$, which corresponds to  the Planck-2018 measurement, $\Omega h^2 = 0.120$~\cite{Planck:2018vyg}, with an assumed 20\% theoretical uncertainty.
	\item {\bf DM direct and indirect detections.} The spin-dependant (SD) and spin-independent (SI) DM-nucleon scattering cross-sections should be lower than their upper limits from the latest XENON-1T experiments~\cite{Aprile:2018dbl,Aprile:2019dbj}. In addition, the prediction of the gamma-ray spectrum from DM annihilation in the dwarf spheroidal galaxies should agree with the limit placed by the Fermi-LAT observations~\cite{Fermi-LAT:2015att}. This restriction was implemented by the joint-likelihood analysis suggested in~\cite{Carpenter:2016thc}.
	\item {\bf Higgs physics.} The lightest CP-even Higgs boson corresponds to the SM-like Higgs boson discovered at the LHC. Its properties should coincide with corresponding data obtained by ATLAS and CMS collaborations at 95\% confidence level. This requirement was examined by program \texttt{HiggsSignal-2.2.3}~\cite{HS2013xfa,HSConstraining2013hwa,HS2014ewa,HS2020uwn}. In addition, the direct searches for extra Higgs bosons at LEP, Tevatron and LHC were checked by program \texttt{HiggsBounds-5.3.2}~\cite{HB2008jh,HB2011sb,HB2013wla,HB2020pkv}.
	\item {\bf B-physics.} The branching ratios of $B_s \to \mu^+ \mu^-$ and $B \to X_s \gamma$ should agree with their experimental measurements at $2\sigma$ level~\cite{PhysRevD.98.030001}.
\item {\bf Vacuum stability.} The vacuum state of the scalar potential consisting of the Higgs fields and the
last two generation slepton fields should be either stable or long-lived. This condition was checked by program Vevacious~\cite{Camargo-Molina:2013qva}.
\end{itemize}

\begin{figure}[t]
	\centering
		\includegraphics[width=0.45\textwidth]{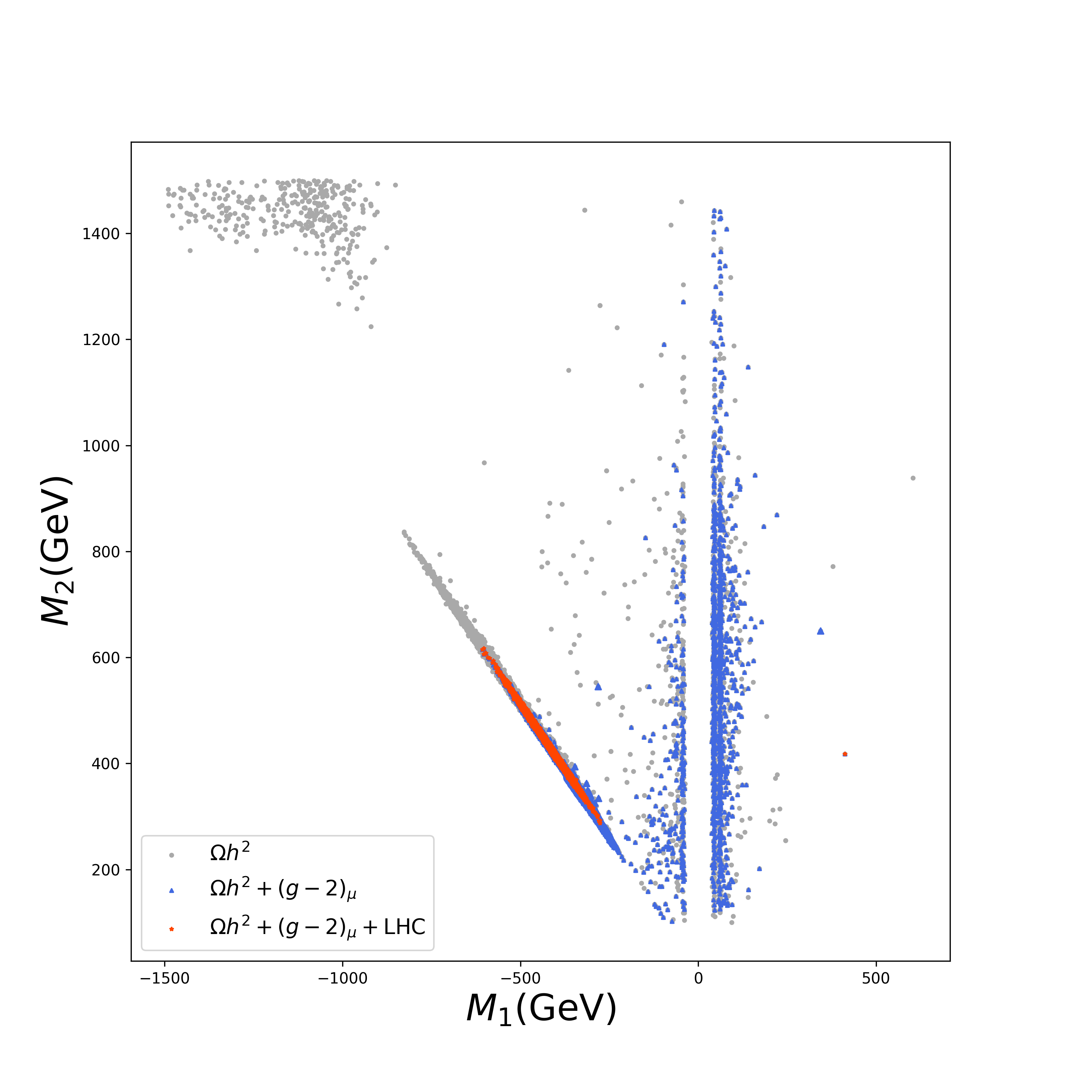}\hspace{-0.3cm}
		\includegraphics[width=0.45\textwidth]{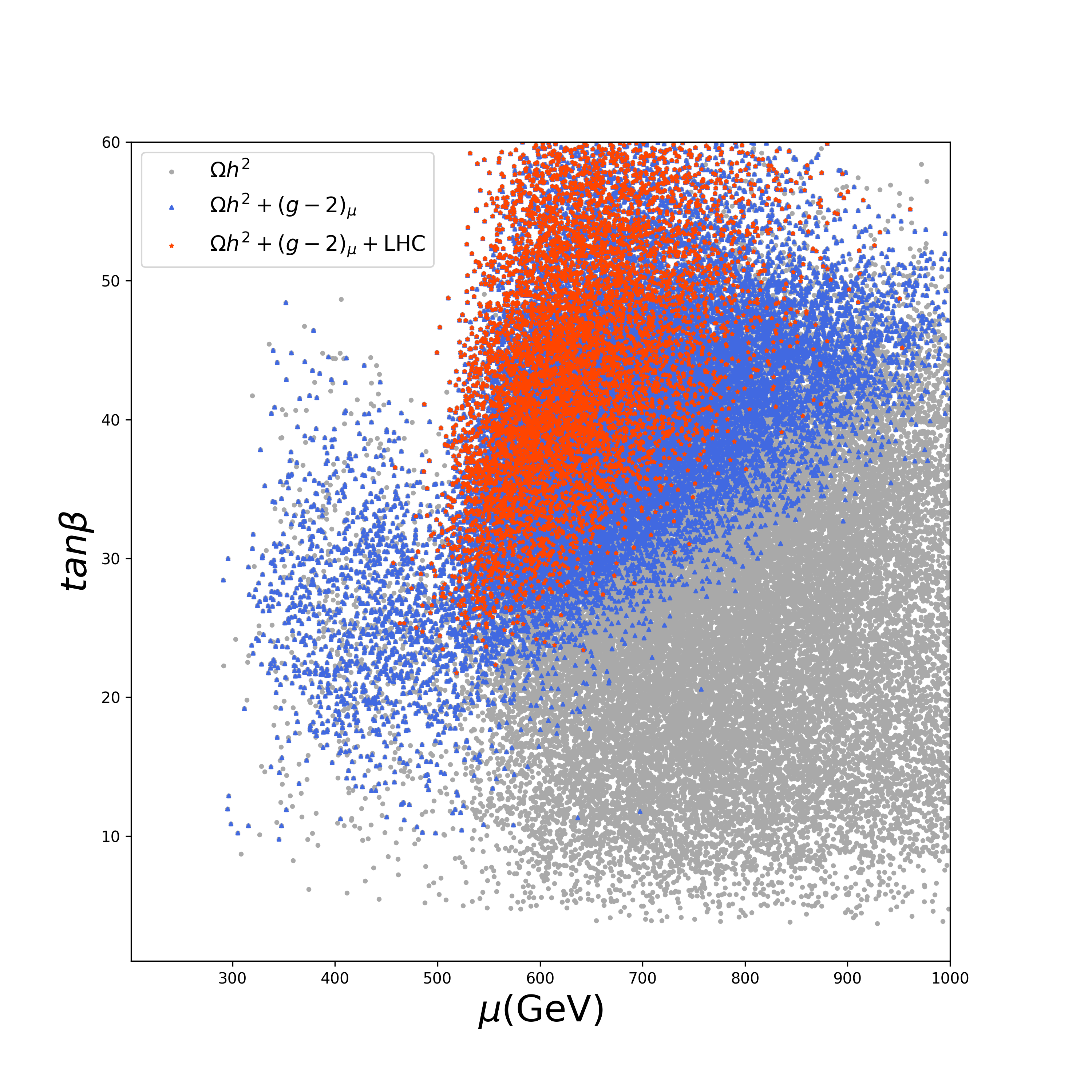}
	\caption{\label{Z3Amufig1} Samples obtained from the parameter scan, which are projected onto the $M_1-M_2$ plane~(left panel) and $\mu-\rm{tan\beta}$ plane~(right panel).
The grey points represent the samples that are consistent with the results of DM physics experiments, the blue triangles denote the ones that can further explain the anomaly at $2\sigma$ level, and the red stars are those that satisfy all experimental constraints, in particular, the limit from the LHC searches for SUSY.}
\end{figure}

To get to know the impact of LHC searches for SUSY on the scan results, the following processes were studied by Monte Carlo simulations:
\begin{equation}\begin{split}
pp \to \tilde{\chi}_i^0\tilde{\chi}_j^{\pm} &, \quad i = 2, 3, 4, 5; \quad j = 1, 2 \\
pp \to \tilde{\chi}_i^{\pm}\tilde{\chi}_j^{\mp} &, \quad i,j = 1, 2; \\
pp \to \tilde{\chi}_i^{0}\tilde{\chi}_j^{0} &, \quad i,j = 2, 3, 4, 5; \\
pp \to \tilde{\mu}_i \tilde{\mu}_j &,\quad i,j = L, R;
\end{split}\end{equation}
Specifically, in order to save computing time, program \texttt{SModelS-2.1.1}~\cite{Khosa:2020zar}, which contains the experimental analyses in Table~\ref{tab:1},  was first used to exclude the obtained samples.
Given this program's capability in implementing the LHC constraints was limited by its database and the strict prerequisites to use it, the rest samples were further surveyed by simulating the analyses listed in Table~\ref{tab:2}. In this research, the cross-sections for each process were calculated to the next-to leading order by program \texttt{Prospino2}~\cite{Beenakker:1996ed}. 60000 and 40000 events were generated for electroweakino and slepton production processes, respectively, by package \texttt{MadGraph\_aMC@NLO}~\cite{Alwall:2011uj, Conte:2012fm} and their parton shower and hadronization were finished by program \texttt{PYTHIA8}~\cite{Sjostrand:2014zea}. Detector simulations were implemented with program \texttt{Delphes}~\cite{deFavereau:2013fsa}. Finally, the event files were put into the package \texttt{CheckMATE\-2.0.29}~\cite{Drees:2013wra,Dercks:2016npn, Kim:2015wza} to calculate the $R$ value defined by $R \equiv max\{S_i/S_{i,obs}^{95}\}$ for all the involved analyses, where $S_i$ represents the simulated event number of the $i$-th signal region (SR), and $S_{i,obs}^{95}$ is
the corresponding $95\%$ confidence level upper limit. Evidently, $R > 1 $ indicates that the sample is experimentally excluded if the involved uncertainties are neglected~\cite{Cao:2021tuh}, while $R < 1$ means that it is consistent with the experimental analyses.

\begin{figure}[t]
	\centering
		\includegraphics[width=0.45\textwidth]{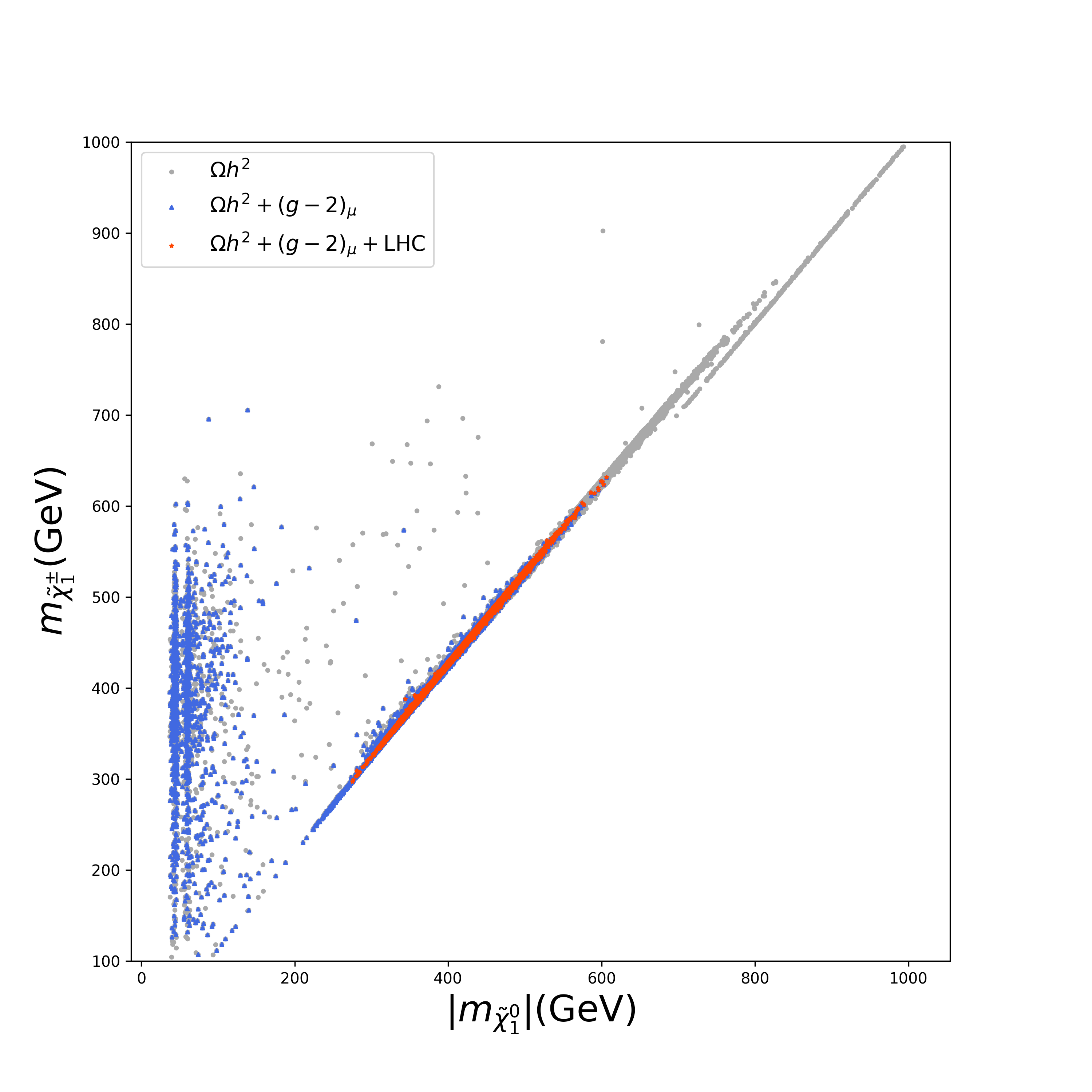}\hspace{-0.3cm}
		\includegraphics[width=0.45\textwidth]{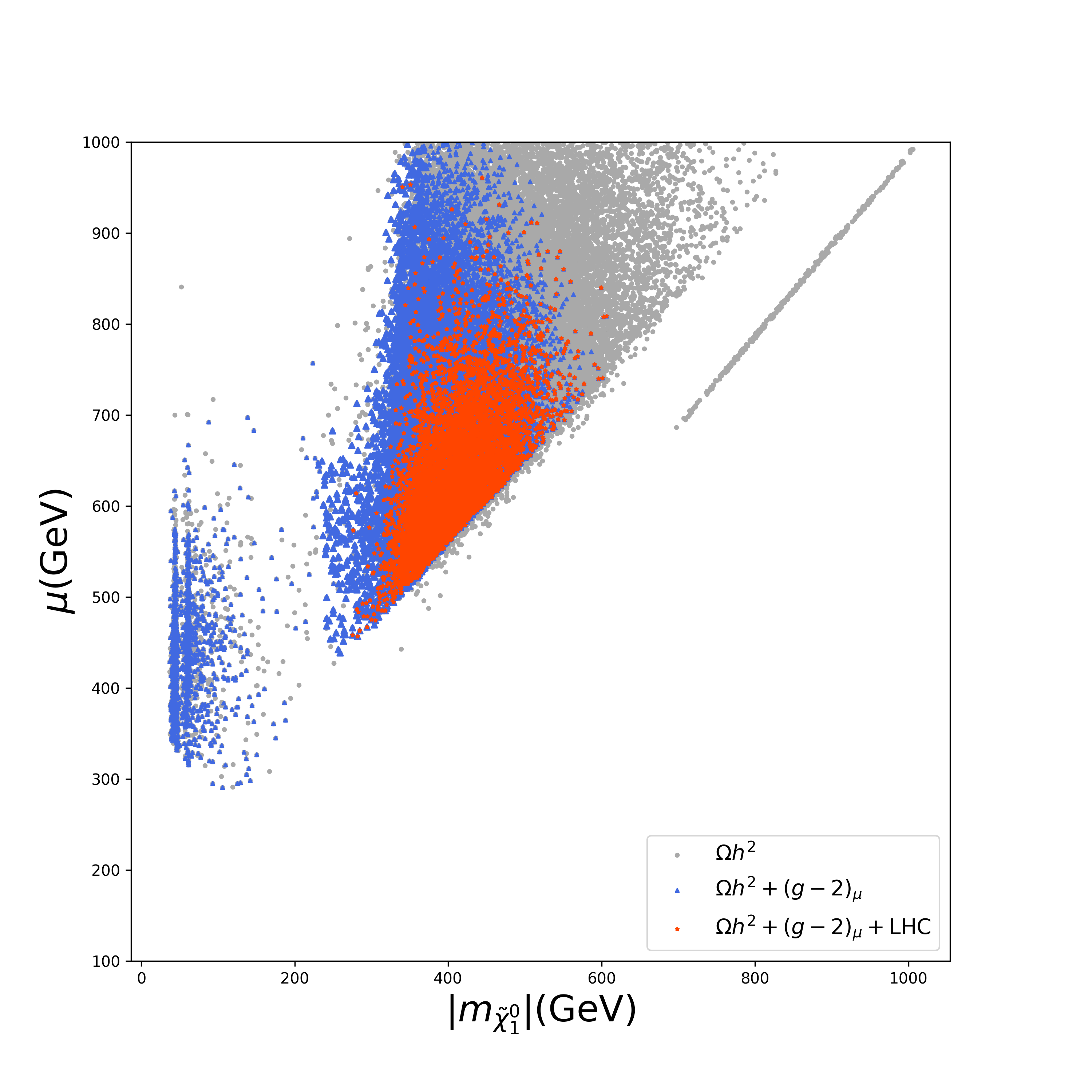}
	\caption{\label{Z3Amufig2}
		Similar to Fig.~\ref{Z3Amufig1}, but projected onto the $|m_{\tilde{\chi}^0_1}|-m_{\tilde{\chi}^\pm_1}$ plane~(left panel) and $|m_{\tilde{\chi}^0_1}|-\mu$ plane~(right panel).}
\end{figure}

\begin{figure}[t]
	\centering
	\includegraphics[width=0.45\textwidth]{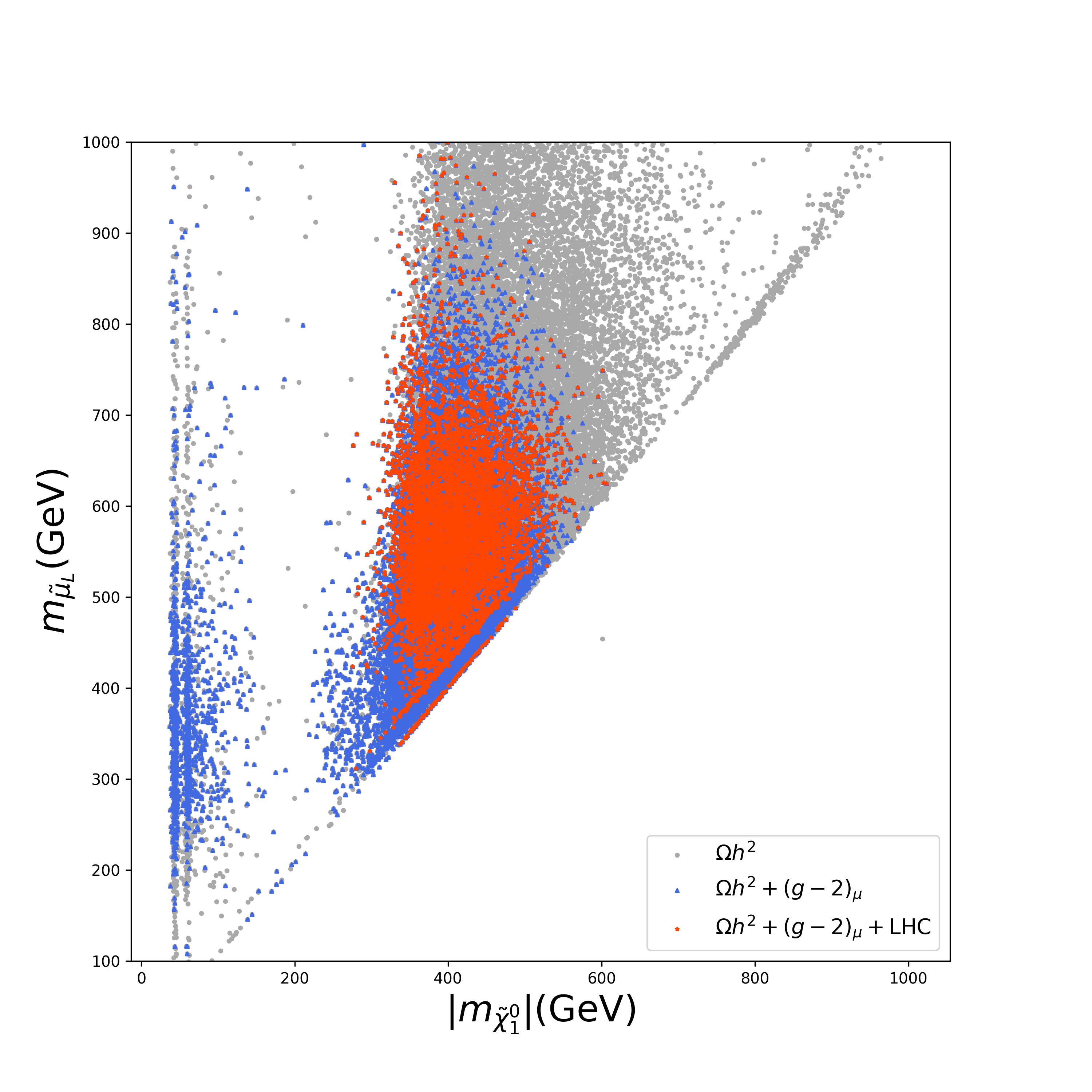}\hspace{-0.3cm}
	\includegraphics[width=0.45\textwidth]{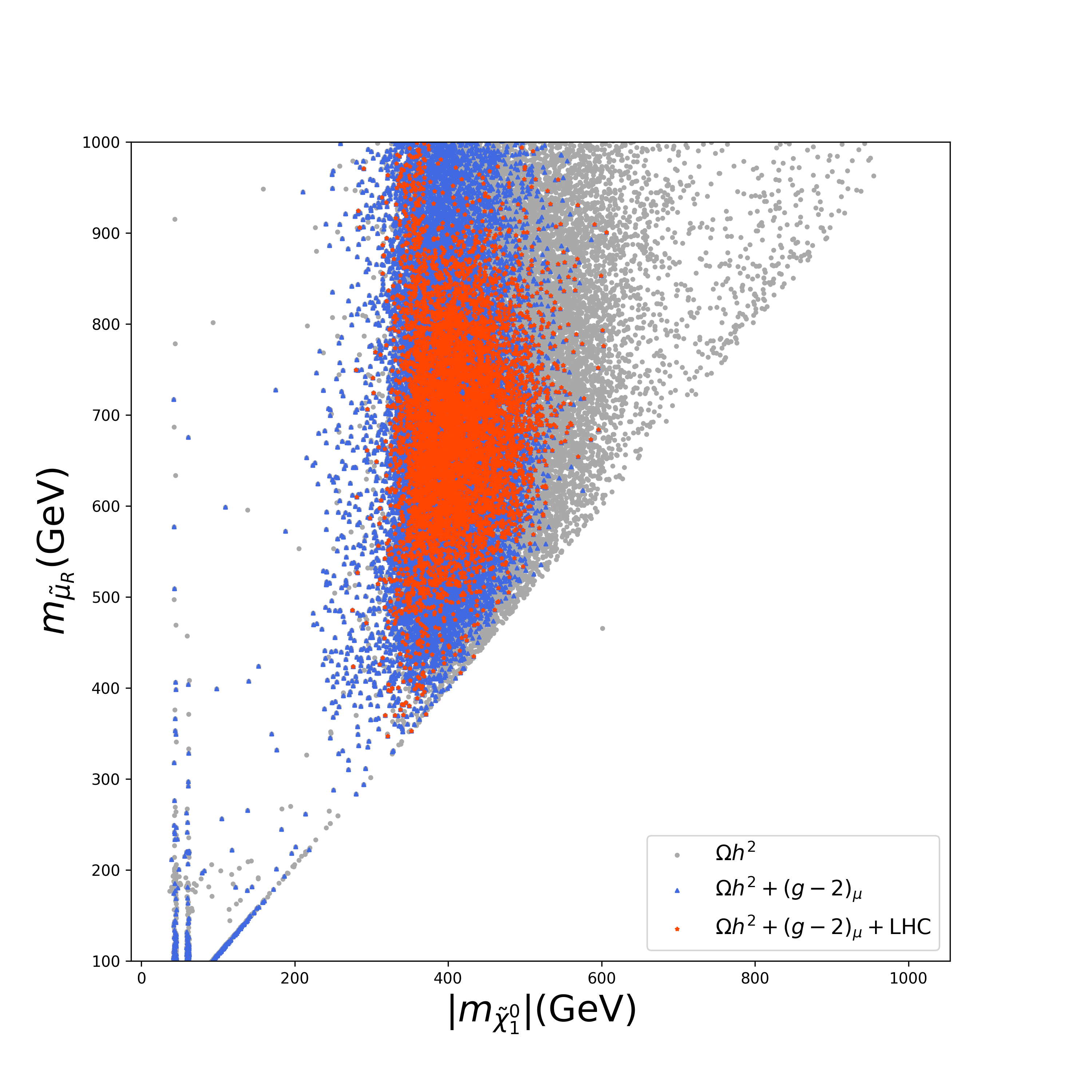}
	\caption{\label{Z3Amufig3}
		Similar to Fig.~\ref{Z3Amufig1} and Fig.~\ref{Z3Amufig2}, but projected onto the $|m_{\tilde{\chi}^0_1}|-m_{\tilde{\mu}_{\rm{L}}}$ plane (left panel) and $|m_{\tilde{\chi}^0_1}|-m_{\tilde{\mu}_{\rm{R}}}$ plane (right panel).}
\end{figure}

\vspace{-0.2cm}
			
\subsection{\label{region} Key features of the results}

\begin{table}[]
	\centering
	\caption{\label{tab:number} Sample numbers before and after implementing the LHC restrictions. These samples were marked by blue and red color, respectively, in Figs.~\ref{Z3Amufig1}-\ref{Z3Amufig3}. }

\vspace{0.2cm}

	\begin{tabular}{l|c|c}
		\hline
		Annihilation Mechanisms & Without LHC Constraints &  With LHC Constraints       \\ \hline
		\multicolumn{1}{l|}{Total Sample}                                                                          & 21241             & 7280         \\
		\multicolumn{1}{l|}{Bino-Wino Co-annihilation}                                                                    & 18517              & 7189        \\
		\multicolumn{1}{l|}{Bino-Smuon Co-annihilation}                                                                                                   & 1886                 & 87    \\
		\multicolumn{1}{l|}{$Z$-funnel}                                                                    & 401              &  0        \\
		\multicolumn{1}{l|}{$h_1$-funnel}                                                                    & 323              & 0         \\\hline
		
	\end{tabular}
\end{table}

All samples obtained by the scan were projected onto two-dimensional planes in Figs.~\ref{Z3Amufig1}-\ref{Z3Amufig3}, where they were classified by three different colors to distinguish the impacts of the DM experiments, the muon anomaly, and the LHC probes of SUSY on the parameter space. From these plots, the following conclusions can be inferred:
\begin{itemize}
	\item If only the constraints from DM physics are implemented, $\tilde{\chi}_1^0$ is Bino-dominated when $|m_{\tilde{\chi}_1^0}| \lesssim 700~{\rm GeV}$. It may achieve the correct relic density by Z-funnel, $h_1$-funnel, or co-annihilating with Wino-like electroweakinos and/or Smuons. $\tilde{\chi}_1^0$ may also be Higgsino-dominated when $ 800~{\rm GeV} \lesssim |m_{\tilde{\chi}_1^0}| < 1~{\rm TeV}$ and $800~{\rm GeV} \lesssim |M_1| \leq 1.5~{\rm TeV}$. This is a scenario specific to the $\mathbb{Z}_3$-NMSSM~\cite{Cao:2016nix} since the mass splittings among Higgsino-dominated electroweakinos, i.e., $\tilde{\chi}_1^0$, $\tilde{\chi}_2^0$, and $\tilde{\chi}_1^\pm$, depend on $\lambda$ and the Singlino mass $m_{\tilde{S}} \equiv 2\kappa \mu/\lambda$ (see formulae 3.3 in~\cite{Cao:2021lmj}), and consequently the effective cross-section of their co-annihilation may differ sizably from that of the MSSM~\cite{Griest:1990kh,Baker:2015qna}. In the intermediate mass range $700~{\rm GeV} < |m_{\tilde{\chi}_1^0} | < 800~{\rm GeV}$, $\tilde{\chi}_1^0$ may be either Bino-dominated or Higgsino-dominated, which is reflected by the left panel of Fig.~\ref{Z3Amufig2}. In addition, the DM DD experiments have required $\mu \gtrsim 300~{\rm GeV}$, which was explained by analytic formulae in~\cite{Cao:2019qng}.

  \item If the $\mathbb{Z}_3$-NMSSM is further required to explain the $(g-2)_\mu$ anomaly at $2\sigma$ level, $\tilde{\chi}_1^0$ should be lighter than about $620~{\rm GeV}$ and thus Bino-dominated. With the increase of $|m_{\tilde{\chi}_1^0}|$, $\mu$, $m_{\tilde{\mu}_L}$, and  $m_{\tilde{\mu}_R}$ prefer smaller and smaller values. This tendency is more obvious for $\mu$ and $m_{\tilde{\mu}_L}$ than for  $m_{\tilde{\mu}_R}$. The fundamental reason comes from the fact that the $\mathbb{Z}_3$-NMSSM is a decoupled theory in heavy sparticles limit and $\Delta a_\mu^{\rm SUSY}$ is more sensitive to $\mu$ and  $m_{\tilde{\mu}_L}$ than to  $m_{\tilde{\mu}_R}$, which is shown in Eqs. \ref{eq:WHL}-\ref{eq:BLR}. In addition, $\tan \beta$ must be larger than about 10 to explain the anomaly.

  \item The LHC searches for SUSY have significant impacts on the explanation of the $(g-2)_\mu$ anomaly. Specifically, the involved sparticles are set at a lower bound in mass, i.e., $|m_{\tilde{\chi}_1^0}| \gtrsim 275~{\rm GeV}$, $m_{\tilde{\chi}_1^\pm} \gtrsim 300~{\rm GeV}$, $m_{\tilde{\mu}_L} \gtrsim 310~{\rm GeV}$, $m_{\tilde{\mu}_R} \gtrsim 350~{\rm GeV}$, and $\mu \gtrsim 460~{\rm GeV}$. The basic reasons are as follows: if $\tilde{\chi}_1^0$ is lighter, more missing momentum will be emitted in the sparticle production processes at the LHC, which can improve the sensitivities of the experimental analyses; if the sparticles other than $\tilde{\chi}_1^0$ are lighter, they will be more copiously produced at the LHC to increase the events containing multiple leptons. In Appendix A of~\cite{Cao:2022chy}, the reason why $\mu \lesssim 500~{\rm GeV}$ is disfavored was also explained by analytic expressions. Since lighter sparticles are forbidden, $\tan \beta$ has to be larger than about 20 to solve the discrepancy.

\end{itemize}

In Table \ref{tab:number}, the impact of the LHC constraints on DM physics was shown. This table indicates that the resonant annihilations have been completely excluded, and it is the co-annihilation with Wino-like electroweakinos (in most cases) and/or Smuons (for a few cases) that are responsible for the measured relic density\footnote{For the same mass of the electroweakinos and Smuons, the cross-sections of DM co-annihilation with Smuons are much lower than that with the electroweakinos, which causes that Smuons will be much lighter than the electroweakinos to achieve the correct relic density.}. The LHC constraints are more efficient in excluding the Smuon co-annihilating mechanism than the electroweakino co-annihilating mechanism. The reason is that the Smuon corresponds to the NLSP for the former mechanism, and it can increase the leptonic signal rate since heavy sparticles will decay into the Smuon~\cite{Cao:2021tuh,Cao:2022chy}. It was verified that the signal regions for more than 3 leptons of CMS-SUS-16-039 and for more than 200 GeV of $\rm{E}_{\rm{T}}^{\rm{miss}}$ of CMS-SUS-20-001 played a crucial role in excluding the samples.

Concerning the obtained results, several comments are in order:
\begin{itemize}
\item It is evident that the parameter distributions shown in Figs. \ref{Z3Amufig1}-\ref{Z3Amufig3} depend on the scan strategy. To make the conclusions of this study as robust as possible, different strategies, e.g., narrowing or broadening the parameter space in Eq.~(\ref{amuNMSSM-scan}) and/or changing the prior distribution of the inputs, were adopted to compare the obtained results. It was found that the main conclusions were scarcely changed by the strategy. In this aspect, it should be noted that a lower bound on Wino mass, $|M_2| \gtrsim 230~{\rm GeV}$, has been set for the Bino-Wino co-annihilation case by the experimental analyses in~\cite{ATLAS:2021moa}. This conclusion can be directly applied to this study.
\item Analyzing the properties of the samples indicates that all the singlet-dominated particles, including the Singlino-dominated neutralino and CP-even and -odd Higgs bosons, are heavier than $600~{\rm GeV}$. It also indicates that $\lambda < 0.4$ and the Singlino-dominated neutralino never co-annihilates with the Bino-dominated DM to obtain the measured density~\cite{Baum:2017enm}. As a result, the $\mathbb{Z}_3$-NMSSM and MSSM have roughly the same underlying physics in explaining the anomaly, which means that the conclusions in this work should be applied to the MSSM\footnote{Comparing the MSSM results presented in \cite{Chakraborti:2020vjp,Chakraborti:2021dli}, this work reveals at least three common key features. First, the Bino-dominated DM achieves the correct relic density by co-annihilating either with the Wino-dominated electroweakinos or with the Sleptons. Second, the LSP and NLSP are upper bounded by about $600~{\rm GeV}$ in mass, which sets clear search targets for future colliders. Last, the Higgsinos should be heavier than about $500~{\rm GeV}$ to satisfy various experimental constraints and meanwhile predict a sizable $a_\mu^{\rm SUSY}$. }. In addition, this study did not find the case that $\tilde{\chi}_1^0$ was Singlino-dominated. The main reason comes from its suppressed Bayesian evidence~\cite{Zhou:2021pit}. This conclusion was also commented in~\cite{Cao:2022chy}.
\item It is notable that the SUSY explanation of the anomaly will be explored at future colliders since some of the involved sparticles can not be excessively heavy, particularly the LSP and NLSP should be lighter than about $600~{\rm GeV}$. This issue was discussed in Refs.~\cite{Chakraborti:2020vjp,Chakraborti:2021mbr,Chakraborti:2021squ,Chakraborti:2021dli}.
    It was found that, although only a part of the preferred parameter space can be covered at the high luminosity LHC, exhaustive coverage of the parameter space is reachable at a high-energy $e^+ e^-$ collider with $\sqrt{s} \gtrsim 1~{\rm TeV}$, such as ILC with $\sqrt{s} = 1~{\rm TeV}$~\cite{Baer:2013cma} and CLIC with $\sqrt{s} = 1~{\rm TeV}$~\cite{CLICDetector:2013tfe,CLICdp:2018cto}. This conclusion was shown in Fig.4 of~\cite{Chakraborti:2021squ}, where the capability of different colliders to probe the explanation was compared for the Bino-Wino co-annihilation case.
\item Throughout this study, both the theoretical uncertainties incurred by the simulations and the experimental (systematic and statistic) uncertainties were not taken into account. These effects can relax the LHC constraints. However, given the advent of high-luminosity LHC, it is expected that much tighter constraints on the $\mathbb{Z}_3$-NMSSM will be obtained in near future.
\item In some high energy SUSY-breaking theories, $\tilde{\tau}$ may be the NLSP. In this case, the production rate of the $e/\mu$ final states will be changed in comparison with current study. As a result, both the LHC constraints and subsequently the explanation of the anomaly show different features (see, e.g., the discussion in~\cite{Hagiwara:2017lse}). Such a possibility will be discussed in our future work.
\end{itemize}

\vspace{-0.2cm}

\section{\label{conclusion-section}Summary}
The discrepancy between $a_\mu^{\rm Exp}$ and $a_\mu^{\rm SM}$ were recently corroborated by the E989 experiment at FNAL. It hints the existence of new physics, and supersymmetry as the most compelling one has attracted a lot of attention. However, most of the studies focused on the MSSM in explaining the anomaly and few works were carried out in the $\mathbb{Z}_3$-NMSSM. This fact motivates us to explore the implications of the anomaly to this extended theory.

As had long been expected, the $\mathbb{Z}_3$-NMSSM could explain the anomaly in its broad parameter space, and this, in turn, placed special demands on the theory. For example, the LSP and NLSP should be lighter than about $620~{\rm GeV}$ and $650~{\rm GeV}$, respectively, since heavier mass spectra will suppress $a_\mu^{\rm SUSY}$ so that the theory fails to account for the anomaly at $2\sigma$ level. One remarkable improvement of this study over the previous ones, in particular Refs. \cite{Chakraborti:2020vjp} and \cite{Baum:2021qzx}, is that the constraints from the LHC probes of SUSY are surveyed comprehensively to limit the parameter space of the $\mathbb{Z}_3$-NMSSM. As a result, lower bounds on sparticle mass spectra are obtained, i.e.,  $|M_1| \gtrsim 275~{\rm GeV}$, $M_2 \gtrsim 300~{\rm GeV}$,  $\mu \gtrsim 460~{\rm GeV}$, $m_{\tilde{\mu}_L} \gtrsim 310~{\rm GeV}$, and $m_{\tilde{\mu}_R} \gtrsim 350~{\rm GeV}$. The basic reasons for the results are as follows: if $\tilde{\chi}_1^0$ is lighter, more missing momentum will be emitted in the sparticle production processes at the LHC, which can improve the sensitivities of the experimental analyses; while if the sparticles other than $\tilde{\chi}_1^0$ are lighter, they will be more copiously produced at the LHC to increase the events containing multiple leptons. These bounds are far beyond the reach of the LEP experiments in searching for SUSY, and have not been noticed before. They have significant impacts on DM physics, e.g., the popular $Z$- and Higgs-funnel regions have been excluded if the theory is required to explain the $(g-2)_\mu$ anomaly, and the Bino-dominated neutralino DM has to co-annihilate with the Wino-dominated electroweakinos (in most cases) and/or Smuons (in few cases) to obtain the correct density. Furthermore, it is inferred that these conclusions should apply to the MSSM since the underlying physics for the bounds are the same.  This research provides useful information for future SUSY searches at colliders.

\vspace{-0.3cm}	
		
\section*{Acknowledgement}
The authors thank Lei Meng for the helpful discussion about the posterior probability distribution function of the scan performed in this study. This work is supported by the National Natural Science Foundation of China (NNSFC) under grant No. 12075076.
		
\bibliographystyle{CitationStyle}
\bibliography{g-2}

\end{document}